# Advanced Surface Chemistry Analysis of Carbon Nanotube Fibers by X-ray Photoelectron Spectroscopy

*Belén Alemán\*, María Vila and Juan J. Vilatela\**

Dr. B. Alemán, Dr. M. Vila, Dr. J.J. Vilatela

IMDEA Materials Institute, Eric Kandel 2, Getafe, Madrid 28906, Spain

E-mail: belen.aleman@imdea.org, juanjose.vilatela@imdea.org

Carbon nanotube fibers are materials with an exceptional combination of properties, including higher toughness than carbon fibers, electrical conductivity above metals, large specific surface area (250 $m^2$/g) and high electrochemical stability. As such, they are a key component in various multifunctional structures combining augmented mechanical properties with efficient interfacial energy storage/transfer processes. This work presents a thorough XPS study of CNT fibers subjected to different purification and chemical treatments, including spatially-resolved micro XPS synchrotron measurements. The dominant feature is an inherently high degree of $sp^2$ conjugation, leading to a strong plasmonic band and a semi-metallic valence band lineshape. This high degree of CNT perfection in terms of longitudinal "graphitization" helps to explain reported bulk properties including the high electrical and thermal conductivity, and accessible quantum capacitance. There is also presence of organic impurities, mostly heavy carbonaceous molecules formed as by-products during fiber synthesis and which are adsorbed on the CNTs. Sulfur, a promoter used in the CNT growth reaction, is found both in these surface impurities and associated with the Fe catalyst. The observation of strongly-adsorbed surface impurities is consistent with the high ductility of CNT fibers, attributed to interfacial lubricity.

1. Introduction

Surfaces and interfaces are dominant features of nanostructured materials. In the case of nanocarbons, CNTs or graphene of few layers, their surface chemistry has a strong impact on intra and inter-particle properties, including charge transport, stress transfer[1] and catalytic activity.[2] It is equally important for understanding the interaction of nanocarbons with other materials when combined in hybrids or composites.[3,4] In this regard, macroscopic fibers of CNTs (CNTF) are a very interesting system of study, particularly those produced by the direct spinning of a CNT aerogel from the gas-phase during chemical vapor deposition (CVD).[5] Reaction[6] and assembly[7] conditions in this process lead to a macroscopic fiber made up of extremely long CNTs (1mm) with a high degree of perfection and which are associate in long coherent domains (bundles). This leads to efficient exploitation of CNT molecular properties in the bulk, with reports demonstrating superior mechanical to high-performance fibers,[8] and electrical and thermal conductivity similar or above Cu.[9] In the quest to further improvements in longitudinal fiber properties, particularly mechanical, there is consensus that the CNT surface chemistry plays a dominant role in stress transfer, modulating shear strength and modulus between CNTs over several orders of magnitude, affecting deformation both in the elastic and plastic regimes. But although tailoring CNT surface chemistry has become one of the most promising methods to modify bulk CNT fiber tensile properties,[10] the surface chemistry of the starting material is poorly defined. Similarly, there is enormous interest in the use of CNT fiber materials in a wide range of energy storage devices. Their combination of high surface area (250 $m^2/g$), high electrical conductivity (3.5 x $10^5$ S/m) and mechanical toughness (fracture energy above 80 J/g) makes them ideal electrodes, often as both active material and current collector.[11] Amongst multifunctional devices, CNT fibers have been used in all-solid[12–15] and structural[16] supercapacitors, electrocatalyst support/current collector,[17] battery current collectors[18,19] and electrodes,[20] and in a range of photoelectrochemical conversion devices,[21] including counter electrodes in dye-sensitized solar cells[22] with superior efficiency than Pt (9.5% vs 9%).[23] In parallel, it has also been observed that the high quality and few layers of the CNTs lead to bulk electrodes with accessible quantum (chemical) capacitance, reminiscent of 1D confinement in the constituent CNTs and closely related to the electrode joint density of states. [24] However, this vast body of work contrast with the lack of knowledge about the surface chemistry of the CNT fibers, which can be expected to affect direct electrochemical processes, charge transfer from other species, transport properties[25,26] and in general the electronic structure of the CNT fibers. Inspired by previous X-ray photoelectron spectroscopy (XPS) measurements on CNT fibers to analyze doping,[20,27] oxidative chemical functionalization[14] and interfacing with semiconductors, [28] this work sets out to study in detail the surface chemistry of CNT fibers, both as-made and after purification. It focuses on quantitatively determining compositional and plasmonic features in the spectra, analyzing the valence band lineshape and identifying the chemical state and location of residual promotor as a method to clarify its role in the CVD growth stage.

2. Experimental

2.1. Synthesis of CNT fiber

CNT fibers are synthesized by the direct spinning of carbon nanotubes[5] from the gas phase by floating catalyst CVD. Butanol is used as carbon source, ferrocene as catalyst source (Fe) and thiophene as promoter source (S). The reaction is carried out in hydrogen atmosphere at 1250 °C in a vertical furnace, where the mm long CNTs form an elastic aerogel that is drawn and directly spun into a bobbin. All as-made CNT fibers in this work are made of few-layer CNTs[6] with an initial solution composition of 0.8 wt.% of ferrocene, 1.5 wt.% of thiophene and 97.7 wt. % of butanol. In *section 3.2* and *3.3*, fibers with differerent predominant CNT type in terms of nanotube diameter and number of layers (from single-wall (SWNTs) to multi-wall (MWNTs) nanotubes), varying thiophene initial wt. % from 0.1 to 1,[6] are also analyzed. All CNT fibers are spun in non-oriented conditions[7] at high precursor feed rate (~ 5 ml/min) and low winding rate (~ 3 m/min).

2.2. Treatments

For the surface analysis of this work, CNT fiber was put through different treatments. For *section 3.1* and *3.2*, CNT fiber was annealed at 450°C for 12 hours in high vacuum conditions ($10^{-7}$-$10^{-8}$ mbar). In *section 3.3*, on one hand the fiber was kept in ultra-high vacuum (UHV) for degassing time enough to be measured at $10^{-9}$-$10^{-10}$ mbar and on the other hand it was heated up to 1100°C under water vapor atmosphere for 10 min in a tubular furnace where air has been previously evacuated. **Table 1** shows the different CNT fiber samples and conditions for XPS analysis.

Annealing and mild oxidation treatments have the effect of removing surface contaminants from the CNTs. In terms of bulk mechanical properties, the result is a reduction in plastic deformation and an increase in elastic modulus. After annealing in a non-oxidative atmosphere at 450°C, for example, the modulus is roughly doubled and the strain-to-break halved. Mild oxidative functionalization produces similar effects, albeit with a small reduction in tensile strength, most likely on account of the degradation of CNT bundles after introduction of functional groups. Similarly, both types of treatment lead to a decrease in longitudinal electrical conductivity after evaporation of adsorbed water, a known p-dopant of nanocarbons.

**Table 1.** CNT fiber samples composition and conditions for XPS analysis.

| Section | Samples | Composition | Treatment |
|---|---|---|---|
| Section 3.1 | As-made CNTF | Few layer MWNTs | ----- |
| | Annealed CNTF | Few layer MWNTs | 450°C at $10^{-8}$ mbar |
| Section 3.2 | As-made CNTF | Few layer MWNTs | ----- |
| | Annealed CNTF | Few layer MWNTs | 450°C at $10^{-8}$ mbar |
| | CNTFs with different CNT type | SWNTs- MWNTs | Sulfur variation in synthesis |
| Section 3.3 | As-made CNTF | Few layer MWNTs | ----- |
| | Degassed CNTF | Few layer MWNTs | Exhaustive degasification |
| | Water vapor CNTF | Few layer MWNTs | 1100 °C, 10 min in water vapor |
| | CNTFs with different CNT type | SWNTs- MWNTs | Sulfur variation in synthesis |

2.3. CNT fiber characterization

Morphological characterization of the CNT fibers was carried out using a dual beam with field-emission scanning electron microscope FIB-FEGSEM, Helios NanoLab 600i FEI at 15 kV and a transmission electron microscope (TEM) EOL JEM 3000F TEM at 300 kV. Raman spectroscopy was performed by a Renishaw PLC spectrometer with 532 nm wavelength laser excitation.

2.4. XPS characterization

XPS measurements were collected with both monochromatic Al-Kα (hν = 1486.71 eV) and synchrotron radiation X-ray sources using *Phoibos 150 and 100* hemispherical energy analyzers (SPECS GmbH). Synchrotron radiation measurements were carried out at ESCAmicroscopy beamline (ELETTRA synchrotron in Trieste, Italy) using a photon energy of 650 eV in a scanning photoelectron microscope (SPEM), which can perform chemical imaging and micro-spectroscopic analysis with a lateral resolution better than 100 nm using a X-ray microprobe focused to a submicron spot onto the sample. In the imaging mode, photoelectrons are collected with a selected kinetic energy, revealing the distribution of a particular element or elemental chemical state of the surface. All the spectra within a section and analysis have been obtained in the same equipment under the same conditions.

C1s core level spectra were fitted by Gaussian/Lorentzian peak shapes and Shirley background profile was subtracted. The fitting criteria for CNT fiber XPS analysis is based on the selection of four carbon related components at ~284.5 eV (C=C), ~285 eV (C-C, C-H), ~286 (C-O) eV and ~290 eV ($\pi$-$\pi$* loss band) taking into account that high binding energy components (C-C, C-H, C-O and $\pi$-$\pi$*) are broad components (FWHM $\geq$1) with an enhanced Gaussian character compared to the narrower (FWHM <1) graphitic peak (C=C).

3. Results and discussion

**Figure 1a** shows the photograph of a macroscopic CNT fiber bobbin continuously spun for 40 min. The fiber is made of a network of mm-long CNTs packed in bundles (**Figure 1b**) with diameters in the range of 50 nm, which are preferentially oriented along the fiber axis. The characteristic Raman spectrum of a standard CNT fiber made up of few-layer CNTs (**Figure 1c**) shows the presence of the G band at ~1580 cm$^{-1}$, D band at ~1350 cm$^{-1}$ and overtone modes in the range of 2400-3000 cm$^{-1}$.[29] The high graphitic nature (G band) of these CNTs with respect to the amount of defects (D band) as i.e. carbonaceous impurities with sp$^3$ bonding or broken sp$^2$ bonds, is represented by the intensity ratio of D to G bands ($I_D/I_G$), which comes out at around 0.25. Note, however, that in this context "degree of graphitization" refers to the degree of perfection of the CNT hexagonal lattice, corresponding more appropriately to the degree of sp$^2$ conjugation or longitudinal crystallinity. In layered graphitic systems, graphitization implies regular commensurate ABAB stacking, as in Bernal graphite, whereas the arrangement of graphitic layers (inter and intra) of CNTs is turbostratic, even for highly perfect CNTs. Furthermore, the G band is strongly resonant and sensitive to CNT number of layers, being most intense when the excitation laser line matches optical transitions in SWNTs. For reference, fibers made up of SWNTs give a $I_D/I_G \leq 0.1$.

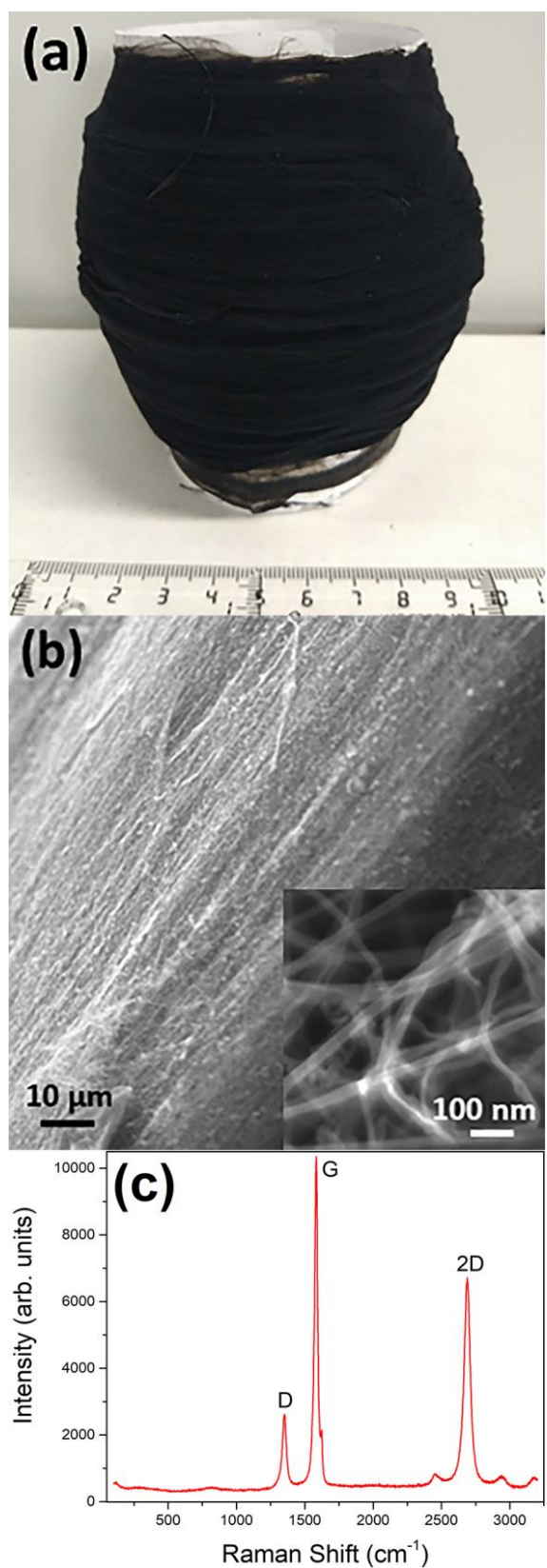

**Figure 1.** (a) Macroscopic photograph of CNTF bobbin made of (b) CNT bundles oriented preferentially along the fiber axis. (c) Raman spectrum of CNTF showing the characteristic

features of few-layer CNTs (G band at ~1580 cm$^{-1}$, D band at ~1350 cm$^{-1}$ and the rest of overtone modes in the range of 2400-3000 cm$^{-1}$).

The aim then, is to use XPS to study the degree of sp$^2$ conjugation in CNT fibers and to identify contributions from surface impurities. The strategy is to analyze samples subjected to different purification methods, and to use spatially-resolved XPS to selectively observe CNT bundles without the presence of neighboring carbonaceous impurities.

3.1. Analysis of purified samples.

One of the first concerns in the surface analysis of carbon materials is to distinguish the chemical composition of the material from the carbonaceous thin layer formed by its exposure to air. This adventitious carbon layer[30] is deposited in all surfaces exposed to the atmosphere and it is made of non-graphitic carbon but polymeric hydrocarbons with a variety of carbon-oxygen species. In addition, CNT fiber contains ~10 wt. % of a carbonaceous layer consisting of by-products of the CVD reaction, mainly as hydroxyl functionalized aliphatic-entities, that adsorb on the CNT fiber at the cold end of the reactor. Most of this carbonaceous layer is volatile and can be desorbed below 400°C. Thus, annealing of CNT fiber in UHV, more conveniently inside the XPS probing chamber, is expected to remove these impurities, as well as adventitious carbon and water molecules physically adsorbed on the CNT surface. Similarly, scanning photoelectron microscopy, which has a spatial resolution up to 100 nm, enables collection of XPS spectra on CNT bundles, thus reducing other contributions from different carbon structures present in the fiber, as low magnification TEM micrograph shows in **Figure 2a**. Higher magnification images confirm that the CNTs are highly regular, as seen in the HRTEM micrograph example in **Figure 2b**, shown as perfectly aligned walls in the CNT and *(hk0)* reflections in the fast Fourier transform. However, some areas of the CNTs and bundles are coated by a carbonaceous layer (**Figure 2c**) and in addition to the CNTs, there is around 7 wt. % of residual encapsulated catalyst, [31] typically in poorly-graphitized tubules (**Figure 2d**).

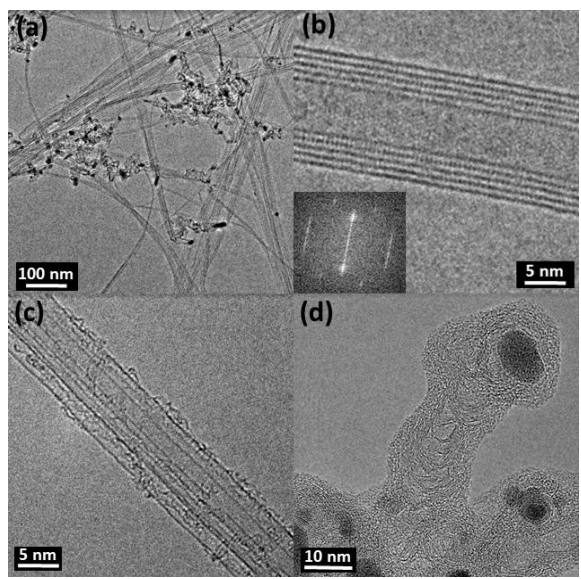

**Figure 2.** TEM images of (a) CNT fiber structure at low magnification and detailed images of the different carbon structures present in the fiber: (b) highly regular graphitized CNT, (c) CNT bundle coated with carbonaceous layer and (d) encapsulated residual catalyst in poorly-graphitized tubules.

We identify the corresponding surfaces under analysis as: as-made, annealed and annealed bundle (**Table 2**).

**Table 2.** Sample conditions to distinguish different C contributions in XPS analysis of CNT fiber.

| Conditions | Adventitious C | Synthesis by-products | Non-tubular structures | CNT bundle |
|---|---|---|---|---|
| As-made CNTF | ✓ | ✓ | ✓ | ✓ |
| Annealed CNTF | ✗ | ✗ | ✓ | ✓ |
| Annealed CNT bundle | ✗ | ✗ | ✗ | ✓ |

**Figure 3** presents the results obtained from these three surfaces. The photoelectron images (**Figure 3a-b**) obtained at C1s binding energy reflect surface topography and confirm that in microprobe measurement conditions, the bundles can be approximately resolved. Analysis of the survey spectra (**Figure 3c**) shows the predominant presence of C1s emission (~ 285 eV), with no major changes after annealing in UHV, expect for a small increase in O1s (~ 530 eV) after impurities desorption and which thus suggests the presence of oxygen chemically bonded to the fiber/bundle/CNTs.

The inset of **Figure 3c** shows the detail region of C1s emission related to secondary interactions (satellite losses) produced by collective electron nature like both π-π* transitions (~290 eV) and plasmon effect due to delocalized π bonds of the graphitic system (~ 314 eV). [32] Annealing produces an enhancement of π-π* transition band, while the plasmon feature remains approximately constant, as expected after the CNT surface has been purified. [33]

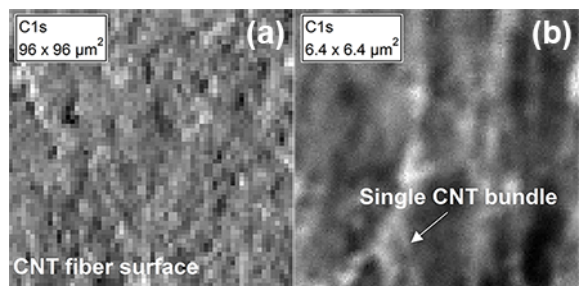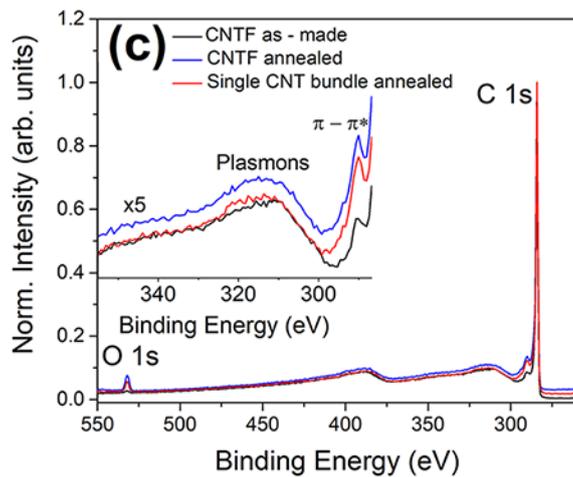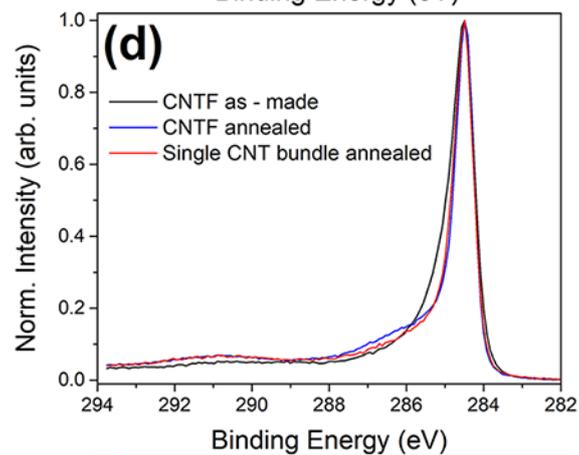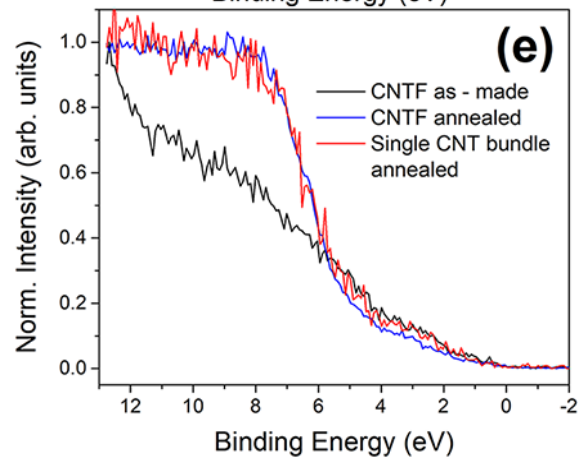

**Figure 3.** Photoelectron images obtained at C1s binding energy of CNTF surface of (a) 96 x 96 μm$^2$ and (b) 6.4 x 6.4 μm$^2$ where a single CNT bundle was measured. (c) Survey spectra normalized at C1s emission showing that the cleaning effect of annealing reveals the presence of oxygen and enhances π-π* transitions in CNTF and CNT bundle. The inset shows a magnification of the high binding energy range of C1s emission where plasmon band (~314 eV) and π-π* transitions (~290 eV) are present. (d) Normalized C1s spectra showing the narrowing effect of annealing in the main peak as well as the π-π* transitions enhancement. (e) Normalized valence band spectra to 13 eV showing the fine structure of CNTs after removing surface impurities.

Closer inspection of the C1s peak shows that it is dominated by the emission of sp$^2$ hybridized C (C=C), i.e., graphitic C, at ~284.5eV (**Figure 3d**). Again, the effect of annealing is observed as a narrowing of the main peak and the enhancement of π-π* loss band at ~290 eV. The contribution from oxygen-containing groups in the range of 286-288 eV can also be better resolved after annealing, which is consistent with the slight increase in O1s in the survey spectra (**Figure 3c**).

Similarly, removal of carbonaceous impurities from the CNT surface reveals its fine structure (**Figure 3e**) in terms of the emissions of C2p π and σ bonds in the ranges of 0-6 eV and 6-12 eV, respectively. [34] The resulting valence band lineshape, with rapidly increasing intensity near 0 eV and an inflection point at ~7.5eV, corresponds to that of a graphitic material with high degree of sp$^2$ conjugation. Although interpretation of valence band emission by XPS is limited due to the low photoionization cross section of C2p core level (0.05 times the cross section of C1s for a photon energy of 650 eV[35]) and the complex nature of these cross sections in the valence band region, [36] the valence band profile in **Figure 3e** is in general agreement with recent measurements on as-made samples analyzed through ultraviolet photoelectron spectroscopy (UPS) with lower photon energy (20-40 eV) and thus enhanced photoemission of C2s and C2p core levels. [28] Nevertheless, the negligible difference in the fine structure between the single bundle and the fiber confirms that bulk surface electronic behavior is representative of the low dimension entities. The comparison of annealed with as-made materials indicates that adsorbed species at the CNT fiber surface probably hybridize with the valence band states due to wave function overlap. This would imply that surface impurities can affect the optical and electronic properties of the constituent CNTs, and therefore bulk properties such as conductance and its dependence on dopants. [25] Such impurities are also expected to have lower electrochemical stability than the CNTs and could thus be the origin of small pseudocapacitive contributions occasionally observed in low scan rate cyclic voltammetry.

3.2. Surface chemistry

The results shown above demonstrate the presence of surface impurities on the constituent CNTs in macroscopic fibers and their effect in broadening and/or masking features of the highly conjugated CNTs underneath. In an effort to relate high-precision XPS measurements to specific surface chemistry features, it is of interest to analyze the different contributions of the C1s core level emission. The fitting assignation method presented (see experimental details) follows extensive XPS measurements on CNT fibers over the last years and represents the best compromise between accuracy and simplicity in number of contributions. Our strategy consists in using four components present in CNT fibers and fixing their initial positions: $sp^2$ hybridized C (C=C) at $284.5\pm0.1$ eV, $sp^3$ hybridized C (C-C, C-H) at $285.0\pm0.5$ eV, C-O groups at $286.0\pm0.5$ eV and $\pi$-$\pi$* loss band at $290\pm1$ eV. [14,37] The resulting spectra are shown in **Figure 4** and the peak details included in **Table 3**.

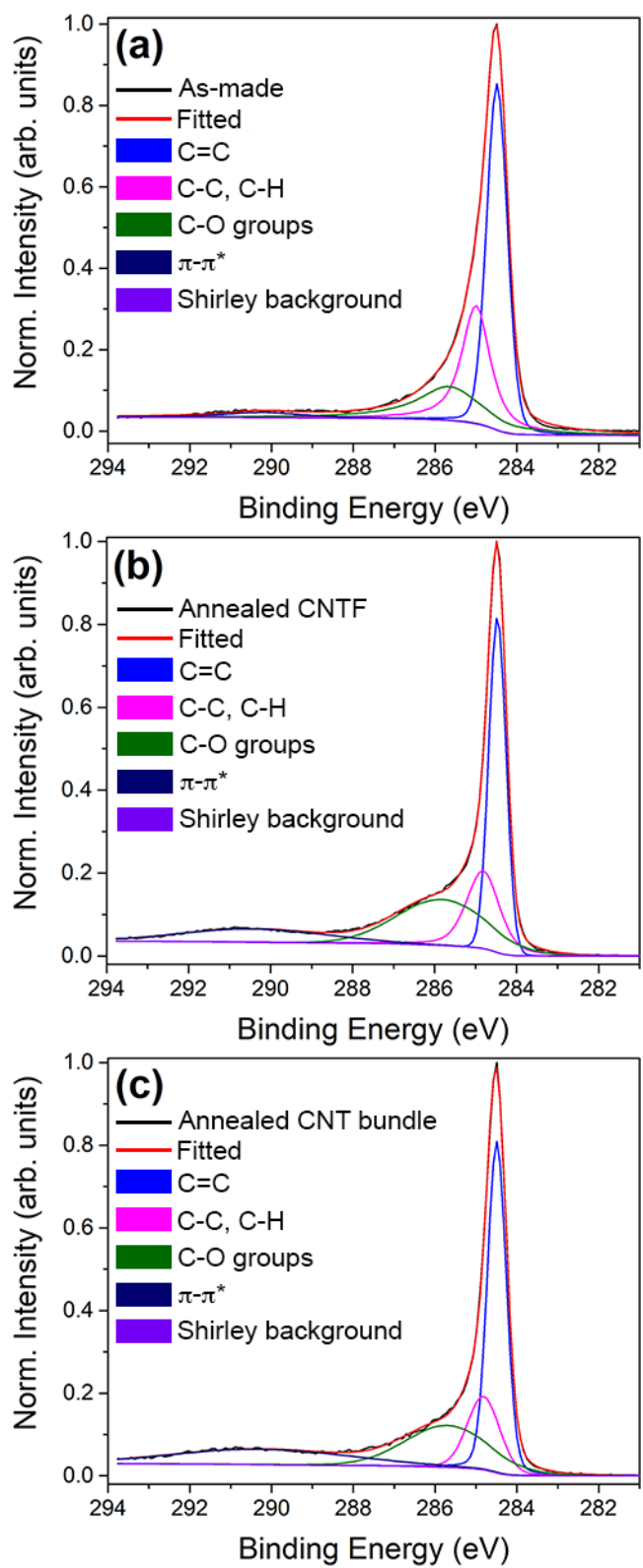

**Figure 4.** C1s core level emission fittings for (a) as-made CNTF, (b) annealed CNTF and (c) annealed single CNT bundle, using components related to sp$^2$ hybridized C (C=C) at 284.5±0.1 eV, sp$^3$ hybridized C (C-C, C-H) at 285.0±0.5 eV, C-O groups at 286.0 ±0.5 eV and π-π* loss band at 290±1 eV.

**Table 3.** Fitting components of C1s spectra for as-made and annealed CNTF, and annealed single CNT bundle.

|  | $sp^2$ hybridized C (C=C) | | $sp^3$ hybridized C (C-C, C-H) | | C-O groups | | $\pi$-$\pi$* | |
| --- | --- | --- | --- | --- | --- | --- | --- | --- |
|  | Peak (eV) | Area (%) | Peak (eV) | Area (%) | Peak (eV) | Area (%) | Peak (eV) | Area (%) |
| **As-made CNTF** | 284.5 | 30 | 284.7 | 34 | 285.5 | 25 | 289.5 | 11 |
| **Annealed CNTF** | 284.5 | 39 | 284.8 | 20 | 285.8 | 28 | 290.4 | 13 |
| **Annealed CNT bundle** | 284.5 | 42 | 284.8 | 16 | 285.7 | 24 | 290.3 | 18 |

From the fitting of the C1s emissions we find that annealing makes individual peaks narrower and better resolved. As expected, removal of impurities leads to a lower contribution $sp^3$-C and an increase in $sp^2$ and $\pi$-$\pi$* peaks, as the surface of CNTs is more exposed after purification. When analyzing the annealed single CNT bundle compared to the extended surface probed on the CNT fiber samples, the bundle shows an enhancement of the graphitic component as well as the $\pi$-$\pi$* transitions, exhibiting the much higher regularity of the CNTs compared with the poorly-graphitized remnant impurity tubules present in the fiber.

Interestingly, after annealing the high binding energy range components related to C-O groups are also better resolved and more intense. They are different from chemical groups purposely introduced via oxidative functionalization, either by gas-phase[14] or electrochemical methods[38], which render the material hydrophilic but eliminate $\pi$-$\pi$* transitions and produce intense features at 288-289 eV, assigned to C=O and C-O=C groups. The observation of this C-O component most likely originates from defects in the hexagonal graphitic structure produced at the point of synthesis as the forming CNT reacts with oxygen species formed from the decomposition of the alcohol precursor used. Further studies on UHV-annealed samples produced from oxygen-free carbon precursors would help clarify this.

A further point of interest is the relatively high ratio of peak areas sp$^3$/sp$^2$, which comes out as 0.4 for the annealed bundle. Clearly, such ratio cannot be equivalent to the number fraction of the particular type of chemical bonds, since topologically a 1mm-long CNT requires a far greater proportion of sp$^2$ bonding. It is possible that the relatively large emission contribution near 285 eV arises partly from bond pyramidalization of the conjugated carbon atoms due to the tube curvature. For 1nm-diameter nanotube the pyramidalization angle ($\theta_p$) is around 4.3°, compared with 0° for a planar graphite and 19.5° for diamond. [39] But for self-collapsed few-layer CNTs of the type found in CNT fibers, [40] the radius of curvature developed at the CNT edges can be very low (~ 0.2 nm), which based on the linear relation between $\theta_p$ and 1/diameter leads to $\theta_p$ as high as 10° and a correspondingly large increase in strain energy (which scales with $(\theta_p)^2$).

In favor of this argument we include in **Figure 5** a comparison of the C1s emission for CNT fibers comprising CNTs of different diameter and number of layers, but similar crystallinity in terms of longitudinal conjugation in the hexagonal lattice according to HRTEM observations. [6]Smaller CNTs clearly lead to a broadening of the sp$^3$ peak and an increase in the sp$^3$/sp$^2$ peak area ratio (from 0.63 for 4.5 nm diameter CNTs to 1.27 for 2 nm diameter CNTs), all consistent with the increase in pyramidalization angle and bond strain energy for smaller or collapsed CNTs.

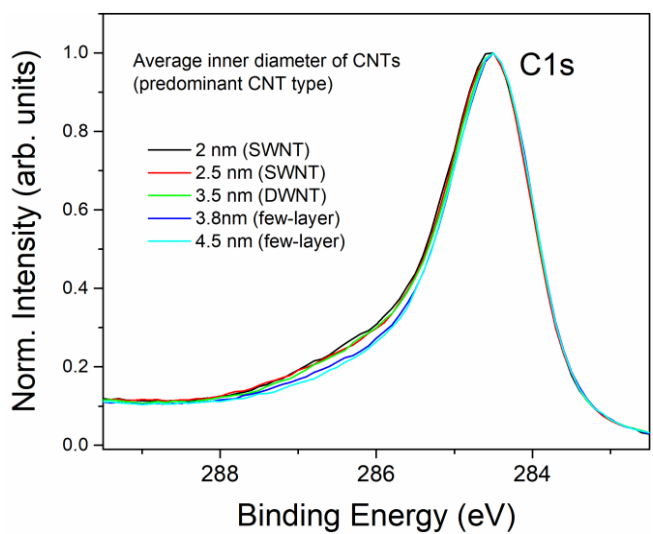

**Figure 5.** C1s core level emission of CNT fibers containing different predominant CNT type, showing broadening of the main peak due to higher sp$^3$ contribution when decreasing the number of walls, consistent with a greater degree of bond pyramidalization due to the higher CNT curvature of small diameter or collapsed CNTs.

3.2. Sulfur identification

The direct CNT fiber spinning process relies on the use of sulfur or another group 16 element as promoter in the floating catalyst CVD reaction. It migrates to the liquid catalyst particle surface, where it limits C solubility in the Fe-rich core, lowers the interfacial energy with the nascent CNT edge while also rejecting its basal plane. It also controls the number of layers of the constituent CNTs through the ratio of S/C. Our previous work using XPS found evidence of sulfur in CNT fiber associated to catalyst particles, observed for example, as Fe-S components in the Fe2p core level emission, in agreement with post-mortem catalyst nanoparticles HRTEM-EDX analysis showing core-shell particles with a Fe-rich core and a S-rich shell. [6,41,42]

To extend our understanding of the location and state of sulfur in CNT fibers, we performed XPS measurements on various samples: as-made, subjected to degasification in UHV ($10^{-9}$-$10^{-10}$ mbar) of the CNT fiber to desorb synthesis by-products, and after oxidation of residual catalyst. The S2p region of the spectra for these samples is included in **Figure 6**, showing two clear broad components at 169 and 164 eV in the as-made sample, generally assigned to $SO_4^{2-}$ and $S^{2-}$ compounds, respectively. The fact that contribution at 169 eV becomes negligible after degasification implies that the adsorbed CVD by-products discussed above contain sulfur. Although not noted before for CNTs, these compounds are common in S-promoted C-films produced by CVD. [43]

Next, we analyze the sample subjected to a mildly oxidizing water vapor treatment. As observed by comparison of the Fe2p region (**Figure 6b**), the degassed CNT fiber catalyst particles present the characteristic metallic Fe comprising a Fe2p doublet (Fe2p$_{3/2}$: 707.1 eV and Fe2p$_{1/2}$: 720.3 eV), broadened in the high energy range by the presence of Fe-S compounds (Fe2p$_{3/2}$: 712 eV and Fe2p$_{1/2}$: 725 eV). After treatment the Fe2p corresponds to that of Fe oxides, accompanied by the appearance of oxide-related emission in O1s region at 530 eV (**inset Figure 6.b**). More importantly, oxidation of the catalyst goes in hand with suppression of S2p components at 162-164 eV.

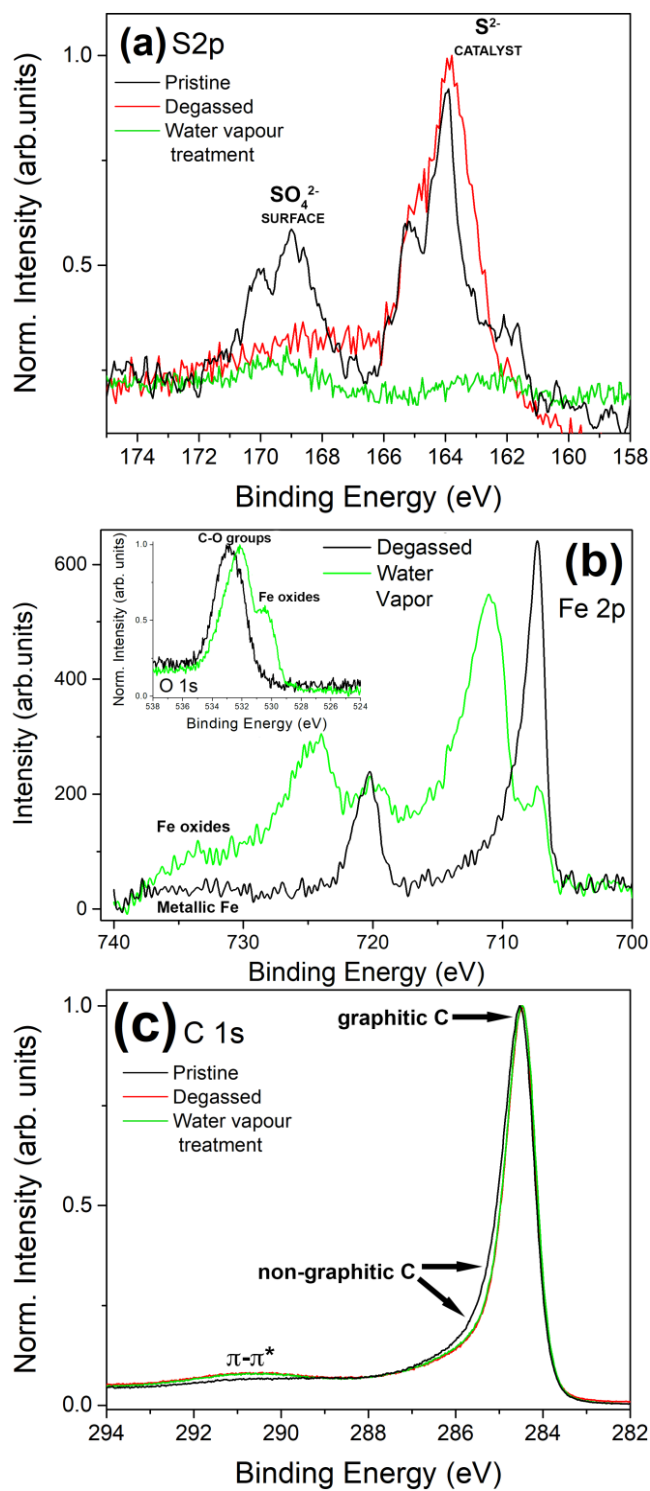

**Figure 6.** (a) S2p core level spectra of as-made, degassed and water vapor treated CNTFs showing the removal of surface related sulfur (169 eV) by the treatments as well as the disappearance of the catalyst related sulfur (162-164 eV) after catalyst oxidation. (b) Fe2p emission showing the oxidation of the catalyst by the water vapor treatment by the increase of Fe-compounds contributions and confirmed by the appearance of O1s component at 530 eV

(inset). (c) C1s emission showing the narrowing of the main peak and enhancement of π-π* transitions band after degassing and water vapor treated CNTF.

Analysis of the C1s spectra further shows that degassing and the water vapor treatment have an effect on "cleaning" the CNT fiber surface, similar to that discussed in above, observed as a reduction in features from non-graphitic carbon at 285 eV while enhancing the π-π* transitions band at 290 eV. The fact that this is observed for a degassing treatment at room temperature confirms that sulfur is not present as a dopant in the hexagonal lattice, which would confer it far higher stability.[44,45] After these observations we can confidently assignation contributions of the S2p peak to organic molecules adsorbed as surface impurities (169 eV) and in the catalyst (162-164 eV), associated to Fe as pyrite ($FeS_2$) or pyhrrotite (FeS),[46,47] respectively . We thus find that in the synthesis of fibers with CNTs with more layers and larger diameter, the ratio of catalyst/surface S decreases rapidly from 0.7 to 0.1 (**Figure 7**). Since we have not observed a correlation between the mass fraction of surface impurities with CNT type, [6] the decrease in catalyst-associated S simply reflects the higher reaction conversion when growing CNTs with more layers, which has the effect of diluting the XPS signal from residual catalyst.

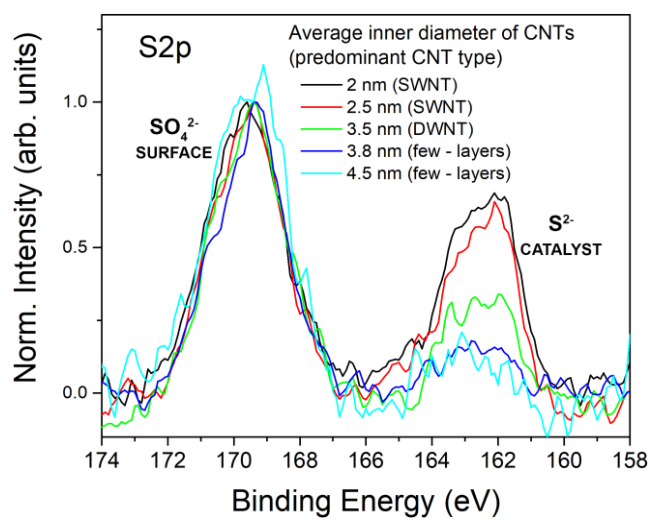

**Figure 7.** XPS spectra of S2p core level for CNT fibers containing different predominant CNT type, produced with increasing sulfur concentrations in the initial reaction. The ratio between catalyst related (164 eV) and surface related (169 eV) sulfur decreases from 0.7 to 0.1 due to the higher number of layers surrounding the catalyst particles which makes the photoelectrons ejected by them less likely to reach the XPS analyzer.

Finally, equipped with the extensive information on CNT surface chemistry discussed above, in **Figure 8** we provide a schematic representation of the most probable surface groups present in as-made and purified CNT fibers, including: adventitious carbon and synthesis by-products in the form of hydrocarbons ($(CH)_n$), oxidized sulfur compounds ($CSO_x$) and oxygen functional groups (-OH, C=O, O=C-OH). Its main purpose is to emphasize that actual CNT fiber materials have very different surface composition from widespread idealized molecular structures, a distinction that can be of particular relevance when describing electrochemical processes, stress transfer and other bulk properties.

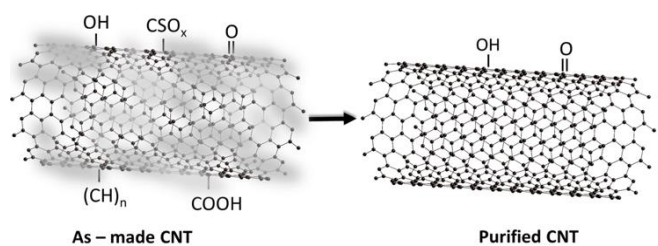

**Figure 8.** Schematic representation of the surface of CNTs in macroscopic CNT fibers produced by the direct spinning method, as-made and after purification.

4. Conclusions

This work presents an advanced study by XPS of CNT fiber subjected to different purification methods. Our results confirm the well-known high graphitic degree of the as-made fiber and the presence of other species at the CNT that may affect its performance in the different applications. Surface impurity removal in CNT fiber by annealing in UHV reveals the high degree of sp$^2$ conjugation by the presence of plasmonic band together with the enhancement of the π- π* transition band in the C1s emission and the fine structure of C2p emission (σ and π bonds) lineshape revealed in the semi-metallic valence band. Purified single CNT bundle, analyzed by spatially resolved XPS, shows the higher regularity of its constituent CNTs compared with other poorly graphitized structures present in the fiber, still in terms of surface analysis, CNT fiber exploits macroscopically the low dimensional behavior of CNTs. The effect of bond piramidalization due to the intrinsic curvature of CNT is also discussed as a contribution with non-graphitic character to C1s emission. Finally, we also identified the presence of residual sulfur promoter not only in the Fe catalyst particles but also as oxidized groups adsorbed at the fiber surface during CVD reaction.

Acknowledgements

The authors are grateful to L. Cabana for the water vapor treatment and to ESCAmicroscopy beamline staff (Elettra synchrotron, Italy) for SPEM experiments. Generous financial support provided by the European Union Seventh Framework Program under grant agreement 678565 (ERC-STEM), is acknowledged. B. Alemán and M. Vila acknowledge Professor Javier Piqueras for his guidance and scientific support through their PhD studies.

Received: ((will be filled in by the editorial staff))
Revised: ((will be filled in by the editorial staff))
Published online: ((will be filled in by the editorial staff))